\documentclass[aps,prl,superscriptaddress,twocolumn,longbibliography]{revtex4}
\usepackage[pdftex]{graphicx}
\usepackage{ulem}
\usepackage{amsmath,amssymb}
\usepackage{color}

\newcommand{\degree}{\ensuremath{^\circ}}
\newcommand{\mathnotation}[2]{\newcommand{#1}{\ensuremath{#2}}}

\newcommand{\VBG}{$V_{bg}$}
\newcommand{\VGone}{V_{\mathrm{g1}}}
\newcommand{\VGtwo}{V_{\mathrm{g2}}}
\newcommand{\VGi}{${V_\mathrm{g \mathit{i}}}$}
\newcommand{\DVGi}{$V_{\mathrm{g} \mathit{i}}^{\mathrm{AC}}$}
\mathnotation{\DVGone}{V_{\mathrm{g} 1}^{\mathrm{AC}}}
\mathnotation{\DVGtwo}{V_{\mathrm{g} 2}^{\mathrm{AC}}}
\newcommand{\VGonedc}{${V_{\mathrm{g1}}^{DC}}$}
\newcommand{\VGtwodc}{${V_{\mathrm{g2}}^{DC}}$}
\newcommand{\VGidc}{$V_{\mathrm{g \mathit{i}}}^{DC}$}
\newcommand{\Gin}{\Gamma_{\mathrm{in}}}

\newcommand{\diff[2]}{\frac{\mathrm{d} #1}{\mathrm{d} #2}}
\newcommand{\ECM}{$E_\mathrm{C}$}
\newcommand{\DPHI}{$\varphi$}
\newcommand{\PHIone}{175\degree{}}
\newcommand{\PHItwo}{100\degree{}}

\newcommand{\GLvalue}{$\Gamma_\mathrm{D}$\,=\,670\,MHz}
\newcommand{\GRvalue}{$\Gamma_\mathrm{S}$\,=\,18\,MHz}
\newcommand{\Tvalue}{$t_{12}/h$\,=\,610\,MHz}
\newcommand{\Ginvalue}{$\Gin{}$\,=\,14\,MHz}

\begin{document}

\title{A two-atom electron pump}
\author{B. ROCHE}
\affiliation{SPSMS, UMR-E CEA / UJF-Grenoble 1, INAC, Grenoble, F-38054, France}
\author{R.-P. RIWAR}
\affiliation{Institut f\"ur Theorie der Statistischen Physik, RWTH Aachen University, D-52056 Aachen, Germany and JARA-Fundamentals of Future Information Technology}
\author{B. VOISIN}
\affiliation{SPSMS, UMR-E CEA / UJF-Grenoble 1, INAC, Grenoble, F-38054, France}
\author{E. DUPONT-FERRIER}
\affiliation{SPSMS, UMR-E CEA / UJF-Grenoble 1, INAC, Grenoble, F-38054, France}
\author{R. WACQUEZ}
\affiliation{CEA, LETI, MINATEC Campus, 17 rue des Martyrs, 38054 GRENOBLE Cedex 9, France}
\author{M. VINET}
\affiliation{CEA, LETI, MINATEC Campus, 17 rue des Martyrs, 38054 GRENOBLE Cedex 9, France}
\author{M. SANQUER}
\affiliation{SPSMS, UMR-E CEA / UJF-Grenoble 1, INAC, Grenoble, F-38054, France}
\author{J. SPLETTSTOESSER}
\affiliation{Institut f\"ur Theorie der Statistischen Physik, RWTH Aachen University, D-52056 Aachen, Germany and JARA-Fundamentals of Future Information Technology}
\author{X. JEHL}
\email[]{xavier.jehl@cea.fr}
\affiliation{SPSMS, UMR-E CEA / UJF-Grenoble 1, INAC, Grenoble, F-38054, France}
\maketitle

\textbf{The fabrication of single atom transistors~\cite{Sellier2006,Pierre2010a,Tan2010,Fuechsle2012} paved the way for electronics based on single dopants~\cite{Koenraad2011,Pla2012}. Recently the spectrum of a single dopant was measured electrically by coupling two such devices~\cite{Roche2012a}.
The next step towards promising functionalities for future nanoelectronics~\cite{Morton2011} consists in manipulating a single electron over two dopants. 
Here we demonstrate electron pumping through two phosphorus donors in series implanted in a silicon nanowire. While quantized pumping is achieved in the low frequency adiabatic regime, we observe remarkable features at higher frequency when the charge transfer is limited by the different tunneling rates.
The transitions between quantum states are modeled involving a Landau-Zener transition, allowing to reproduce in detail the characteristic signatures observed in the non-adiabatic regime.}

Adiabatic charge pumping at zero d.c. bias, in contrast to stationary electronic transport at finite bias, results in a net direct current $I = \mathrm{e}f$ when exactly one electron is transferred during a cycle of period $\frac{1}{f}$~(Refs.~\onlinecite{Kouwenhoven1991,Pothier1992,Thouless1983}).
Motivated by metrological applications~\cite{Pekola2012}, there has been a growing interest in the non-adiabatic regime where this period is comparable to the characteristic time which particles spend in the system~\cite{Moskalets2002,Kaestner2008,Pellegrini2011}. All experiments so far were in either one of these regimes but were not subsequently tuned to both. 
The ultra-scaled electron pump presented here, where the usual large dots are replaced by real nanometer-size atomic orbitals, offers this unique opportunity. 

\begin{figure*}[!t]
\begin{center}
\includegraphics[]{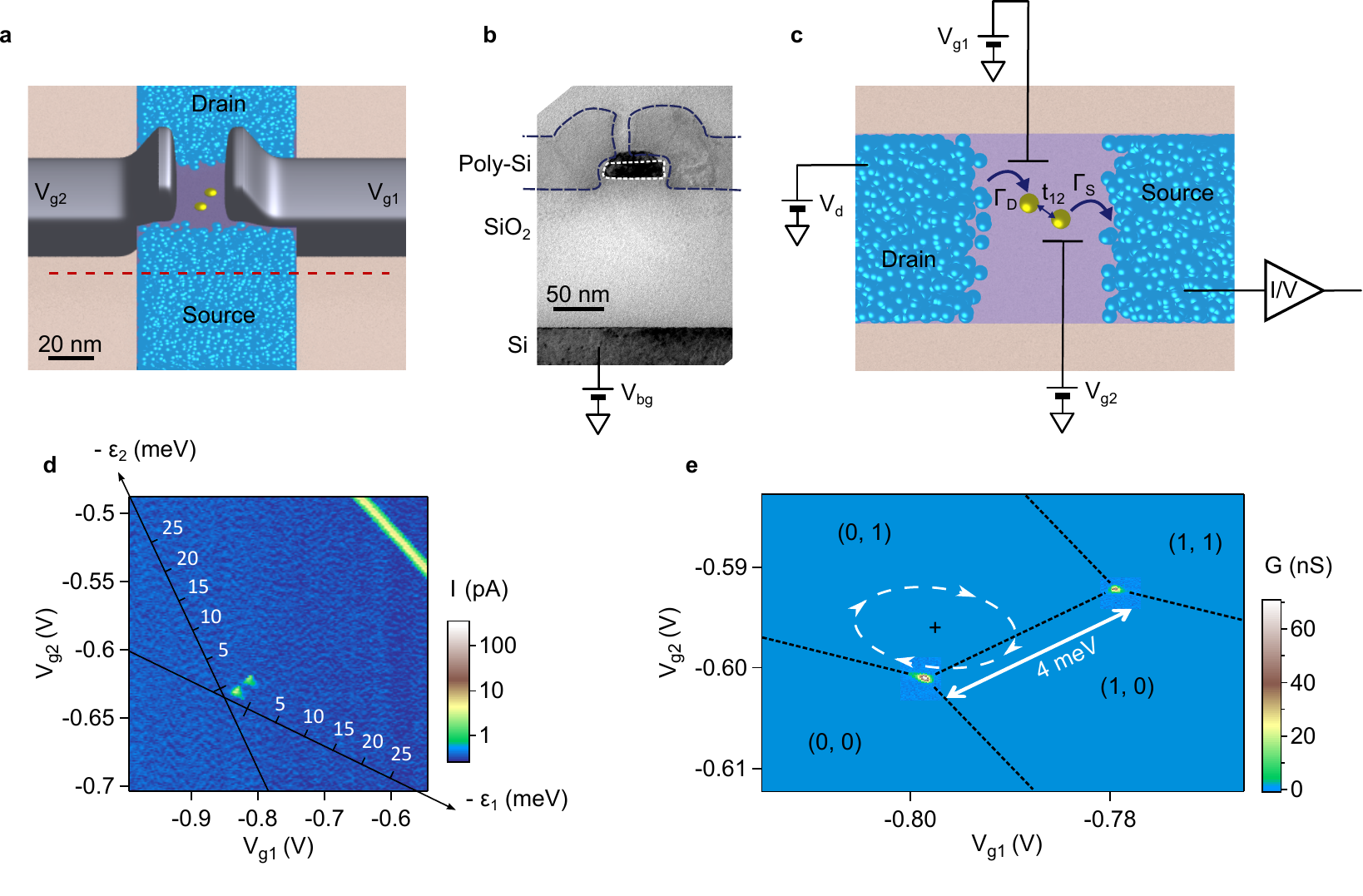}
\caption{\textbf{Control of two dopants with three gates.}
\textbf{a}, Schematic tilted view of the sample along the source-drain axis, with 2 phosphorus dopants in yellow in the small silicon region controlled by 3 gates.
\textbf{b}, Transmission electron micrograph of the cross section of a nominally identical sample. The two top gates and the wire are highlighted by blue and white dashed lines  respectively. The cross section is slightly offset from the gate region (along the red line in \textbf{a}) where the wire is thicker than below the gates because of the source/drain raised by epitaxy. The wafer substrate, separated by a 145\,nm thick SiO$_2$ layer, is used as a back gate.
\textbf{c}, Schematic top view of the sample including the gate and source-drain control parameters.
\textbf{d}, Source-drain current ($I$) with $V_{d}$\,=\,2\,mV, \VBG{}\,=\,$11.5$\,V and superimposed axes giving the energy of each ionized donor with respect to the Fermi level in the leads. The two triangular triple points of the two-dopant system are isolated from any other visible single electron feature by at least 20\,meV.
\textbf{e}, Triple points at small bias ($V_d$\,=\,50\,$\mu$V) measured with a lock-in amplifier. The black dashed lines indicate regions of stable charge states ($n_1$, $n_2$), where $n_i$ is the number of electron on the i$^{th}$ donor. The white dashed line shows an example of pumping trajectory realized in Fig.~\ref{fig3}c, around the working point indicated by the black cross.}
\label{fig1}
\end{center}
\end{figure*}

Single atom transistors are now realized in different ways, including both random~\cite{Tabe2010} or deterministic~\cite{Shinada2005,Fuechsle2012} placement of the dopant atoms.
Electron pumping through many donors has been reported in a larger structure~\cite{Lansbergen2012}. Here, despite the stochastic implantation process, transport through two dopants is obtained because 3 gates control an extremely small volume of silicon (60$\times$20$\times$60\,nm$^3$) in which the number of P dopants is only a few tens for a concentration of $10^{18}$\,cm$^{-3}$, below the metal-insulator transition (see Fig.~\ref{fig1}). The parts of the wire uncovered by the gates and adjacent spacers are doped with arsenic donors at high concentration ($\approx10^{21}$\,cm$^{-3}$) to define conducting source and drain electrodes, separated by 60\,nm. This number prevents transport through a single donor, which requires a much shorter structure~\cite{Pierre2010a}, but favors transport through two donors in series. Indeed a distance of $\approx$~30\,nm between the two donors yields a mean tunnel coupling of the order of 1\,GHz~\cite{Miller1960}, a good compromise between weak coupling and measurable current.
Moreover we obtain separate control over the energy levels of the two dopants by splitting the top gate into two, which face each other at a distance of only 15\,nm.
Finally a voltage \VBG{} is applied to the substrate~\cite{Roche2012a} in order to tune the coupling between the dopants and the leads and rule out rectification effects~\cite{Brouwer2001}, measured for instance in Ref.~\onlinecite{Chorley2012a}. Starting from a nearly perfect yield in terms of transistor behaviour, the final yield for two-dopant features remains higher than 50\% for the geometry and doping level chosen here.

The sample is measured at 150\,mK in a dilution refrigerator. The source-drain current $I$ is first measured with a fixed source-drain bias voltage $V_d$ to select two ideally placed dopants. The back gate voltage is changed until a set of triangles, characteristic of two levels in series~\cite{Wiel2002}, is obtained when scanning the front gates (see Fig.~\ref{fig1}d).
This coupled donor system allows to perform atomic spectroscopy, in contrast with a single donor transistor based on the same technology where fluctuations of the local density of states dramatically affect the current at finite bias~\cite{Vaart1995}. A large (10\,meV) separation in energy between the ground and first excited states was measured in the same sample~\cite{Roche2012a} as investigated here. 
Besides this large valley-orbit splitting which makes it quite insensitive to a small bias, our 2-donor system is also well isolated in energy from any other states which could carry a current from source to drain (see Fig.~\ref{fig1}d). 
The two triple points are separated by a Coulomb repulsion energy \ECM{}\,=\,4.0~$\pm$~0.1\,meV, which is markedly larger than any value reported so far~\cite{Liu2008}.

We now set $V_d$ to zero and modulate the top gates ($i = 1,2$) with phase shifted sinewaves added to the d.c. voltages \VGi{}\,$=$\,\VGidc{} + \DVGi{} sin($2\pi f t \pm \varphi/2$), in order to follow elliptical trajectories in the ($\VGone{}$, $\VGtwo{}$) plane enclosing a triple point~\cite{Pothier1992}(see Fig.~\ref{fig1}e).

\begin{figure}[!t]
\begin{center}
\includegraphics[]{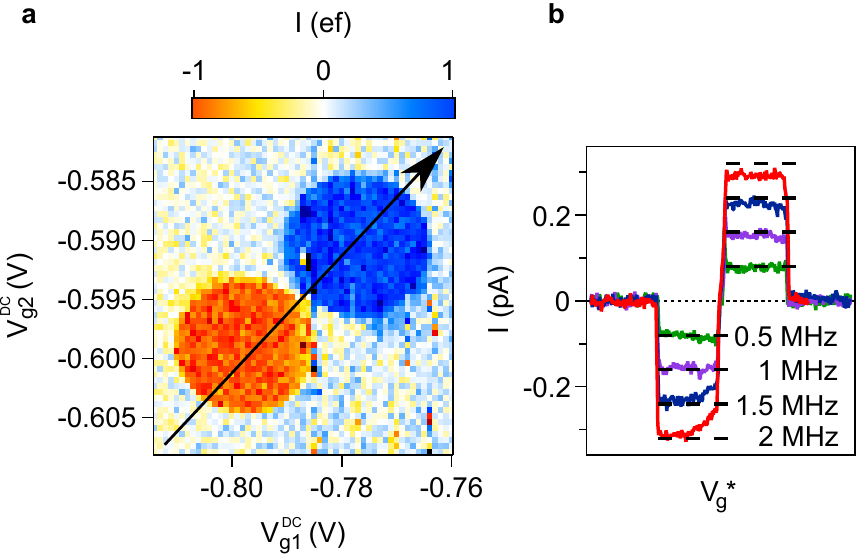}
\caption{\textbf{Adiabatic electron pumping though the two dopants.} \textbf{a}, Map of the pumped current at 1\,MHz and zero bias voltage $V_d$.
\textbf{b}, Pumped current for different frequencies along the $V_g^*$ path indicated by the arrow in \textbf{a}. Quantized pumping $I$\,=\,$\pm$e$f$ is indicated by horizontal dashed lines.}
\label{fig2}
\end{center}
\end{figure}

Adiabatic pumping occurs when the timescale of the driving is much larger than the timescale associated to tunneling to the leads for an electron in the 2-donor system, $\frac{2\pi f v}{\Gamma^2 k_\mathrm{B}T}\ll 1$, and when there are no excitations in the internal dynamics due to the driving signal, requiring {$2\pi t_{12}^2/\hbar v\gg1$ (see Eq.~\ref{Plz}). Here $\Gamma$ is a tunneling rate, $\delta\epsilon = \epsilon_\mathrm{1}-\epsilon_\mathrm{2}$ the difference in energy between the two levels, $v = \left|\diff[\boldsymbol{\delta\epsilon}]{t}\right|$} the speed at which they cross, $k_{B}T$ the energy scale set by the temperature and $t_{12}$ the tunnel coupling between the two levels.
Then the system follows the charge states given by the electrostatic stability diagram (Fig.~\ref{fig1}e). Fig.~\ref{fig2} shows the normalized d.c. current $I/\mathrm{e}f$ measured in this regime. As expected we observe two circular regions of constant quantized current $I = \pm \mathrm{e}f$ wherever a triple point is enclosed by the ellipse, and $I$\,=\,0 elsewhere. The sign of the current is reversed by changing the direction of the trajectory.
The 2-atom system offers intriguing advantages compared to other quantized electron pumps, due to the large energies at play in this system.
The large Coulomb repulsion energy between the dopants ensures wide and flat plateaus for the pumped current versus $V_d$: a slope of 1.9\,G$\Omega$ is obtained over a $V_d$ range of 3.1\,mV at 1\,MHz (see Supplementary Information, Fig.\,S1). 
In contrast with metallic Coulomb islands co-tunneling events via intermediate virtual states are strongly suppressed due to the 10\,meV single level-spacing for each of the dopants~\cite{Roche2012a}. Both these elastic cotunneling effects as well as remaining unwanted events due to inelastic tunneling are shown to be small.

\begin{figure*}[!t]
\begin{center}
\includegraphics[]{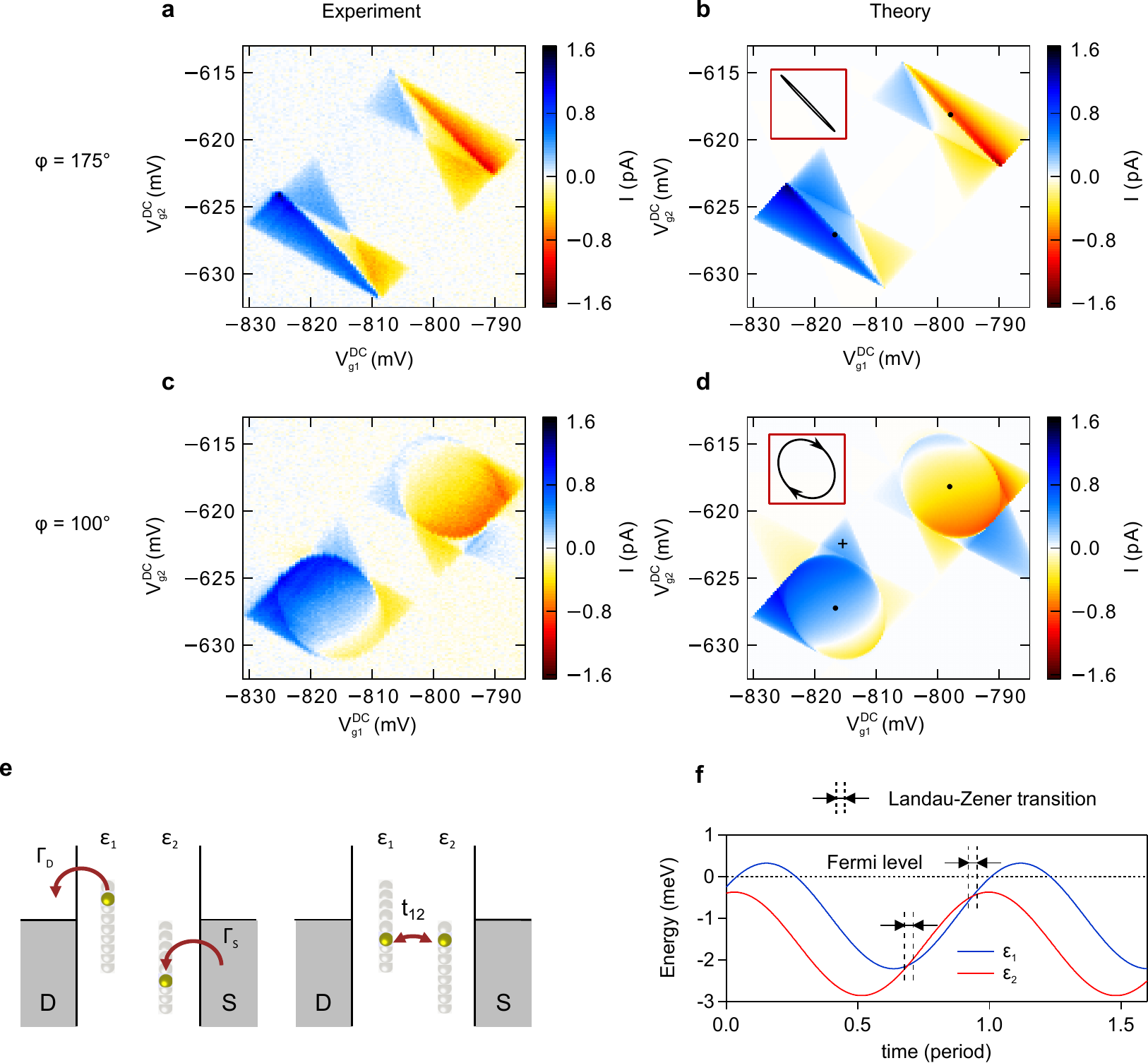}
\caption{\textbf{Non-adiabatic electron pumping at 10~MHz: experiment and simulation.}
\textbf{a},\textbf{c}, 2D map of the current measured with $\varphi$\,=\,175\degree (\textbf{a}) and $\varphi$\,=\,100\degree (\textbf{c}) between the two driving signals as a function of the working point (\VGonedc{}, \VGtwodc{}).
\textbf{b},\textbf{d}, Simulation of the current with \GLvalue{}, \GRvalue{} and \Tvalue{}. The shape of the trajectory in the ($\VGone{}$, $\VGtwo{}$) plane is shown at the top left of the simulation for each phase difference (not to scale), and triple points are indicated with a dot.
\textbf{e}, Sketch of the model used for the previous simulation. Tunneling to the leads occurs with rates $\Gamma_\mathrm{D}$ and $\Gamma_\mathrm{S}$. Coupling between the two dopants by a tunnel amplitude $t_{12}$, leads to a Landau-Zener transition when the two levels cross each other.
\textbf{f}, Calculated energy levels as a function of time for the working point indicated by the black cross in \textbf{d}. Landau-Zener transitions occur as immediate events when the two levels cross (between the dashed lines).}
\label{fig3}
\end{center}
\end{figure*}

At higher frequency the pumped current deviates from the fundamental relation $I$\,=\,e$f$. Figs.~\ref{fig3}a,c show the current map for $f$\,=\,10\,MHz~\footnote{The triple points are slightly shifted as compared to Figs.~\ref{fig1} and \ref{fig2} because the sample was warmed up at room temperature in the meantime.}.
The two circular regions appearing when enclosing the triple points are still visible in Fig.~\ref{fig3}c, but the current inside is not independent of the working point anymore. This is the first manifestation of the deviation from the adiabatic regime. The second remarkable feature is the occurence of extra current outside these circular regions. This effect becomes dramatic in Fig.~\ref{fig3}a, where, due to the vanishingly small surface area of the pumping cycle for \DPHI{}\,=\,\PHIone{}, no signal is expected in the adiabatic case.

The complex features observed in the non-adiabatic regime can be fully understood by means of the model depicted in Fig.~\ref{fig3}e and described in detail in the Supplementary Information. The transitions occuring within the two dopants are described through the dynamics of the closed two-level system. A dissipative part of the dynamics accounts for the tunnel coupling to the source and drain. The transition rates between donor states due to tunneling to the leads are obtained from Fermi's golden rule~\cite{Ingold1992} and are proportional to the tunneling rates to source and drain ($\Gamma_\mathrm{S}$ and $\Gamma_\mathrm{D}$). Hopping of an electron between the two dopants, with tunnel amplitude $t_{12}$, can occur when the two levels cross each other and the resulting time evolution of the two-level system is described according to the Landau-Zener problem~\cite{Zener1932}. Consequently, the probability of staying on the same dopant after the crossing, i.e. the transition from ground to excited state, is given by the Landau-Zener transition probability $P_\text{LZ}$.
Then the probability for an electron to move from one dopant to the other when the levels are aligned depends on the speed $v$ and is given by,
\begin{equation}
1-P_\mathrm{LZ} = 1-\mathrm{exp}\left(-2\pi \frac{t_{12}^2}{\hbar v}\right)
\label{Plz}
\end{equation}
The strict separation of the dynamics of the closed system and of the coupling to the reservoirs is possible because the time during which the levels cross, $\sim 2t_{12}/v$, is of the order of $0.1$ ns and thus much smaller than the time scale related to the dissipative dynamics, $\Gamma_{\text{S}}^{-1},\Gamma_\text{D}^{-1}$.

Using the electrostatic couplings obtained from the d.c. measurements we reproduce the various features observed in the experiment with a unique set of parameters (\GLvalue{}, \GRvalue{} and \Tvalue{}), as illustrated in Figs.~\ref{fig3}b,d for \DPHI{}\,=\,\PHIone{} and \DPHI{}\,=\,\PHItwo{}. The whole set of phase differences \DPHI{} is simulated with the same fidelity (see Supplementary Material, Animation 1). 
If nonadiabatic, the system does not necessarily relax to the ground state, either due to slow tunneling to the leads or due to its internal dynamics ($P_\text{LZ}$ differs significantly from zero, if the condition $2\pi t_{12}^2/\hbar v\gg1$ is no longer fulfilled). Each case yields characteristic features we discuss in the following.

We focus on the working point indicated by the black cross in Fig.~\ref{fig3}d. In this region no current is detected in the adiabatic regime. The time evolution of the energy levels is shown in Fig.~\ref{fig3}f~\footnote{Although the phase shift of the two driving voltages have a phase difference \DPHI{}\,=\,\PHItwo{} the energy level of the two dopants are phase shifted only by 44\degree{} due to the cross-talk between the gates and dopants, see supplementary materials for more details.}.
Dopant 2 can accept electrons only from the drain and not from dopant 1 because it always lies below the Fermi level. After the electron moves to the dopant 1, its energy can be raised above the Fermi level, and it is eventually released into the drain lead. The key point here is that the transition probability between the two discrete dopant levels is different from one, otherwise the electron sitting on dopant 1 would go back to the initial dopant 2 after the second level crossing. This missed tunnel event due to the Landau-Zener transition occuring with probability $P_\text{LZ}\neq 0$ in the non-adiabatic regime increases significantly the error rate~\cite{Keller1996}. Indeed, for this working point, we get $P_{\mathrm{LZ}}\approx$\,0.25 from the simulation for both crossings. In contrast, at low enough frequencies the electron always stays on the lowest energy level according to the adiabatic theorem ($P_{\mathrm{LZ}}\approx 0$). Also the rightmost red non-adiabatic feature can be understood in these terms.

The non adiabatic feature inside the circle is of different origin: here, it is the tunneling dynamics which matters. Even though $\epsilon_\mathrm{2}$ is driven from above to below the Fermi energy before $\epsilon_\mathrm{1}$, the slow tunneling dynamics, $2\pi f v \sim\Gamma^{2}k_{\mathrm{B}}T$, allows to charge dopant 1 first instead of 2. This effect is important in particular if the coupling between dopant 1 and the drain is much stronger than the coupling between dopant 2 and the source. Indeed, $\Gamma_\mathrm{D}\gg\Gamma_\mathrm{S}$ is fulfilled in the experiment, giving rise to the sign change within the circle. Note that this effect would survive even in the limit of $P_{\mathrm{LZ}}\sim0$, hence it does not rely on a non-adiabatic level crossing. The additional features can be understood analogously and are influenced by the individual coupling strengths $\Gamma_\mathrm{S}$ and $\Gamma_\mathrm{D}$, with the exception of one feature that will be discussed in the following.

In order to reproduce the faint side triangle along the anti-diagonal, visible in Fig.~\ref{fig3}d, we finally add a small effective inelastic relaxation rate \Ginvalue{} between levels to the model. However this rate has a very limited impact on the overall pumped current. For the working point indicated by the cross, the simulation shows that only 10\,\% of the total current comes from this inelastic term, the remaining 90\,\% being due to the elastic coupling $t_{12}$. A much higher inelastic rate would suppress the current coming from the Landau-Zener transition.

Importantly, the ultimate atomic size of the system is crucial to get large level separation and strong Coulomb interaction. This enables us to use driving amplitudes corresponding to an energy level modulation of $\approx$~2.5\,meV, much higher than the electronic temperature (300\,mK) in the contact, equivalent to 25\,$\mu$eV. As a consequence the pumped current signal has remarkably sharp borders between different pumping regions, allowing a detailed study of the mechanisms at play in each of these regions. Moreover, due to the large amplitudes, inter-level transitions occur far away (with respect to $k_B T$) from the resonances with the leads, which justifies the zero temperature approach for our model. Finally, since the crucial parameter determining the non-adiabaticity is the speed of the crossing, the actual frequency to observe these effects depends on the size of the ellipse. With the large amplitudes we use the crossover is significantly lowered towards lower frequencies, making it easily accessible experimentally.

We have investigated electron pumping through two dopant atoms connected in series and addressed with three gates. We achieved quantized pumping in the adiabatic regime and, studying the crossover to the non-adiabatic regime, observed additional features in the pumping current. A model taking into account the tunnel couplings to the leads and between the two donors energy levels reproduces all the details of the measured pumped current. This ultimate size system behaves like a nearly perfect two-level system, and the full control of the well separated states in the dopants opens new opportunities for the study of pure spin pumping in a quantum system contacted to ferromagnetic leads~\cite{Riwar2010} and for transporting quantum information~\cite{Greentree2004}.

\begin{acknowledgments}
{\bf \vspace{5mm}Acknowledgements}\\ \scriptsize{We thank V. N. Golovach for fruitful discussions as well as D. Lafond and H. Dansas from CEA-LETI for the transmission electron microscope image.
The authors acknowledge financial support from the EC FP7 FET-proactive NanoICT under Project AFSiD No 214989, the French ANR under Projects SIMPSSON and POESI and from the German Ministry of Innovation NRW.}
\end{acknowledgments}


\begin{thebibliography}{10}
\expandafter\ifx\csname url\endcsname\relax
  \def\url#1{\texttt{#1}}\fi
\expandafter\ifx\csname urlprefix\endcsname\relax\def\urlprefix{URL }\fi
\providecommand{\bibinfo}[2]{#2}
\providecommand{\eprint}[2][]{\url{#2}}

\bibitem{Sellier2006}
\bibinfo{author}{Sellier, H.} \emph{et~al.}
\newblock \bibinfo{title}{Transport spectroscopy of a single dopant in a gated
  silicon nanowire}.
\newblock \emph{\bibinfo{journal}{Phys. Rev. Lett.}}
  \textbf{\bibinfo{volume}{97}}, \bibinfo{pages}{206805}
  (\bibinfo{year}{2006}).
\newblock

\bibitem{Pierre2010a}
\bibinfo{author}{Pierre, M.} \emph{et~al.}
\newblock \bibinfo{title}{Single-donor ionization energies in a nanoscale cmos
  channel}.
\newblock \emph{\bibinfo{journal}{Nature Nanotech.}}
  \textbf{\bibinfo{volume}{5}}, \bibinfo{pages}{133--137}
  (\bibinfo{year}{2010}).

\bibitem{Tan2010}
\bibinfo{author}{Tan, K.~Y.} \emph{et~al.}
\newblock \bibinfo{title}{Transport spectroscopy of single phosphorus donors in
  a silicon nanoscale transistor}.
\newblock \emph{\bibinfo{journal}{Nano Letters}} \textbf{\bibinfo{volume}{10}},
  \bibinfo{pages}{11--15} (\bibinfo{year}{2010}).

\bibitem{Fuechsle2012}
\bibinfo{author}{Fuechsle, M.} \emph{et~al.}
\newblock \bibinfo{title}{A single-atom transistor}.
\newblock \emph{\bibinfo{journal}{Nat Nano}} \textbf{\bibinfo{volume}{7}},
  \bibinfo{pages}{242--246} (\bibinfo{year}{2012}).

\bibitem{Koenraad2011}
\bibinfo{author}{Koenraad, P.~M.} \& \bibinfo{author}{Flatte, M.~E.}
\newblock \bibinfo{title}{Single dopants in semiconductors}.
\newblock \emph{\bibinfo{journal}{Nat Mater}} \textbf{\bibinfo{volume}{10}},
  \bibinfo{pages}{91--100} (\bibinfo{year}{2011}).

\bibitem{Pla2012}
\bibinfo{author}{Pla, J.~J.} \emph{et~al.}
\newblock \bibinfo{title}{A single-atom electron spin qubit in silicon}.
\newblock \emph{\bibinfo{journal}{Nature}} \textbf{\bibinfo{volume}{489}},
  \bibinfo{pages}{541--545} (\bibinfo{year}{2012}).

\bibitem{Roche2012a}
\bibinfo{author}{Roche, B.} \emph{et~al.}
\newblock \bibinfo{title}{Detection of a large valley-orbit splitting in
  silicon with two-donor spectroscopy}.
\newblock \emph{\bibinfo{journal}{Phys. Rev. Lett.}}
  \textbf{\bibinfo{volume}{108}}, \bibinfo{pages}{206812}
  (\bibinfo{year}{2012}).

\bibitem{Morton2011}
\bibinfo{author}{Morton, J. J.~L.}, \bibinfo{author}{McCamey, D.~R.},
  \bibinfo{author}{Eriksson, M.~A.} \& \bibinfo{author}{Lyon, S.~A.}
\newblock \bibinfo{title}{Embracing the quantum limit in silicon computing}.
\newblock \emph{\bibinfo{journal}{Nature}} \textbf{\bibinfo{volume}{479}},
  \bibinfo{pages}{345--353} (\bibinfo{year}{2011}).

\bibitem{Kouwenhoven1991}
\bibinfo{author}{Kouwenhoven, L.~P.}, \bibinfo{author}{Johnson, A.~T.},
  \bibinfo{author}{van~der Vaart, N.~C.}, \bibinfo{author}{Harmans, C. J.
  P.~M.} \& \bibinfo{author}{Foxon, C.~T.}
\newblock \bibinfo{title}{Quantized current in a quantum-dot turnstile using
  oscillating tunnel barriers}.
\newblock \emph{\bibinfo{journal}{Phys. Rev. Lett.}}
  \textbf{\bibinfo{volume}{67}}, \bibinfo{pages}{1626--1629}
  (\bibinfo{year}{1991}).

\bibitem{Pothier1992}
\bibinfo{author}{Pothier, H.}, \bibinfo{author}{Lafarge, P.},
  \bibinfo{author}{Urbina, C.}, \bibinfo{author}{Esteve, D.} \&
  \bibinfo{author}{Devoret, M.~H.}
\newblock \bibinfo{title}{Single-electron pump based on charging effects}.
\newblock \emph{\bibinfo{journal}{EPL (Europhysics Letters)}}
  \textbf{\bibinfo{volume}{17}}, \bibinfo{pages}{249} (\bibinfo{year}{1992}).

\bibitem{Thouless1983}
\bibinfo{author}{Thouless, D.~J.}
\newblock \bibinfo{title}{Quantization of particle transport}.
\newblock \emph{\bibinfo{journal}{Phys. Rev. B}} \textbf{\bibinfo{volume}{27}},
  \bibinfo{pages}{6083--6087} (\bibinfo{year}{1983}).

\bibitem{Pekola2012}
\bibinfo{author}{Pekola, J.~P.} \emph{et~al.}
\newblock \bibinfo{title}{Single-electron current sources: towards a refined
  definition of ampere}.
\newblock \emph{\bibinfo{journal}{arXiv:1208.4030}}  (\bibinfo{year}{2012}).

\bibitem{Moskalets2002}
\bibinfo{author}{Moskalets, M.} \& \bibinfo{author}{B\"uttiker, M.}
\newblock \bibinfo{title}{Floquet scattering theory of quantum pumps}.
\newblock \emph{\bibinfo{journal}{Phys. Rev. B}} \textbf{\bibinfo{volume}{66}},
  \bibinfo{pages}{205320} (\bibinfo{year}{2002}).

\bibitem{Kaestner2008}
\bibinfo{author}{Kaestner, B.} \emph{et~al.}
\newblock \bibinfo{title}{Single-parameter nonadiabatic quantized charge
  pumping}.
\newblock \emph{\bibinfo{journal}{Phys. Rev. B}} \textbf{\bibinfo{volume}{77}},
  \bibinfo{pages}{153301} (\bibinfo{year}{2008}).

\bibitem{Pellegrini2011}
\bibinfo{author}{Pellegrini, F.} \emph{et~al.}
\newblock \bibinfo{title}{Crossover from adiabatic to antiadiabatic quantum
  pumping with dissipation}.
\newblock \emph{\bibinfo{journal}{Phys. Rev. Lett.}}
  \textbf{\bibinfo{volume}{107}}, \bibinfo{pages}{060401}
  (\bibinfo{year}{2011}).

\bibitem{Tabe2010}
\bibinfo{author}{Tabe, M.} \emph{et~al.}
\newblock \bibinfo{title}{Single-electron transport through single dopants in a
  dopant-rich environment}.
\newblock \emph{\bibinfo{journal}{Phys. Rev. Lett.}}
  \textbf{\bibinfo{volume}{105}}, \bibinfo{pages}{016803}
  (\bibinfo{year}{2010}).

\bibitem{Shinada2005}
\bibinfo{author}{Shinada, T.}, \bibinfo{author}{Okamoto, S.},
  \bibinfo{author}{Kobayashi, T.} \& \bibinfo{author}{Ohdomari, I.}
\newblock \bibinfo{title}{Enhancing semiconductor device performance using
  ordered dopant arrays}.
\newblock \emph{\bibinfo{journal}{Nature}} \textbf{\bibinfo{volume}{437}},
  \bibinfo{pages}{1128--1131} (\bibinfo{year}{2005}).

\bibitem{Lansbergen2012}
\bibinfo{author}{Lansbergen, G.~P.}, \bibinfo{author}{Ono, Y.} \&
  \bibinfo{author}{Fujiwara, A.}
\newblock \bibinfo{title}{Donor-based single electron pumps with tunable donor
  binding energy}.
\newblock \emph{\bibinfo{journal}{Nano Letters}} \textbf{\bibinfo{volume}{12}},
  \bibinfo{pages}{763--768} (\bibinfo{year}{2012}).

\bibitem{Miller1960}
\bibinfo{author}{Miller, A.} \& \bibinfo{author}{Abrahams, E.}
\newblock \bibinfo{title}{Impurity conduction at low concentrations}.
\newblock \emph{\bibinfo{journal}{Phys. Rev.}} \textbf{\bibinfo{volume}{120}},
  \bibinfo{pages}{745--755} (\bibinfo{year}{1960}).

\bibitem{Brouwer2001}
\bibinfo{author}{Brouwer, P.~W.}
\newblock \bibinfo{title}{Rectification of displacement currents in an
  adiabatic electron pump}.
\newblock \emph{\bibinfo{journal}{Phys. Rev. B}} \textbf{\bibinfo{volume}{63}},
  \bibinfo{pages}{121303} (\bibinfo{year}{2001}).

\bibitem{Chorley2012a}
\bibinfo{author}{Chorley, S.~J.}, \bibinfo{author}{Frake, J.},
  \bibinfo{author}{Smith, C.~G.}, \bibinfo{author}{Jones, G. A.~C.} \&
  \bibinfo{author}{Buitelaar, M.~R.}
\newblock \bibinfo{title}{Quantized charge pumping through a carbon nanotube
  double quantum dot}.
\newblock \emph{\bibinfo{journal}{Appl. Phys. Lett.}}
  \textbf{\bibinfo{volume}{100}}, \bibinfo{pages}{143104--3}
  (\bibinfo{year}{2012}).

\bibitem{Wiel2002}
\bibinfo{author}{van~der Wiel, W.~G.} \emph{et~al.}
\newblock \bibinfo{title}{Electron transport through double quantum dots}.
\newblock \emph{\bibinfo{journal}{Rev. Mod. Phys.}}
  \textbf{\bibinfo{volume}{75}}, \bibinfo{pages}{1} (\bibinfo{year}{2002}).

\bibitem{Vaart1995}
\bibinfo{author}{van~der Vaart, N.~C.} \emph{et~al.}
\newblock \bibinfo{title}{Resonant tunneling through two discrete energy
  states}.
\newblock \emph{\bibinfo{journal}{Phys. Rev. Lett.}}
  \textbf{\bibinfo{volume}{74}}, \bibinfo{pages}{4702--4705}
  (\bibinfo{year}{1995}).

\bibitem{Liu2008}
\bibinfo{author}{Liu, H.~W.} \emph{et~al.}
\newblock \bibinfo{title}{Pauli-spin-blockade transport through a silicon
  double quantum dot}.
\newblock \emph{\bibinfo{journal}{Phys. Rev. B}} \textbf{\bibinfo{volume}{77}},
  \bibinfo{pages}{073310--} (\bibinfo{year}{2008}).

\bibitem{Ingold1992}
\bibinfo{author}{Ingold, G.-L.} \& \bibinfo{author}{Nazarov, Y.~V.}
\newblock \emph{\bibinfo{title}{Single Charge Tunneling}}, chap. \bibinfo{chapter}{Charge Tunneling
  Rates in Ultrasmall Junctions} (\bibinfo{publisher}{Plenum Press, New York},
  \bibinfo{year}{1992}).

\bibitem{Zener1932}
\bibinfo{author}{Zener, C.}
\newblock \bibinfo{title}{Non-adiabatic crossing of energy levels}.
\newblock \emph{\bibinfo{journal}{Proceedings of the Royal Society of London.
  Series A}} \textbf{\bibinfo{volume}{137}}, \bibinfo{pages}{696--702}
  (\bibinfo{year}{1932}).
\newblock

\bibitem{Keller1996}
\bibinfo{author}{Keller, M.~W.}, \bibinfo{author}{Martinis, J.~M.},
  \bibinfo{author}{Zimmerman, N.~M.} \& \bibinfo{author}{Steinbach, A.~H.}
\newblock \bibinfo{title}{Accuracy of electron counting using a 7-junction
  electron pump}.
\newblock \emph{\bibinfo{journal}{Applied Physics Letters}}
  \textbf{\bibinfo{volume}{69}}, \bibinfo{pages}{1804--1806}
  (\bibinfo{year}{1996}).

\bibitem{Riwar2010}
\bibinfo{author}{Riwar, R.-P.} \& \bibinfo{author}{Splettstoesser, J.}
\newblock \bibinfo{title}{Charge and spin pumping through a double quantum
  dot}.
\newblock \emph{\bibinfo{journal}{Phys. Rev. B}} \textbf{\bibinfo{volume}{82}},
  \bibinfo{pages}{205308} (\bibinfo{year}{2010}).

\bibitem{Greentree2004}
\bibinfo{author}{Greentree, A.~D.}, \bibinfo{author}{Cole, J.~H.},
  \bibinfo{author}{Hamilton, A.~R.} \& \bibinfo{author}{Hollenberg, L. C.~L.}
\newblock \bibinfo{title}{Coherent electronic transfer in quantum dot systems
  using adiabatic passage}.
\newblock \emph{\bibinfo{journal}{Phys. Rev. B}} \textbf{\bibinfo{volume}{70}},
  \bibinfo{pages}{235317} (\bibinfo{year}{2004}).

\end{thebibliography}
\end{document}